\documentstyle[12pt,a4,epsfig]{article}
\begin{document}
\begin{center}
NUCLEON--NUCLEON AND PION--NUCLEON POTENTIALS\\
FROM PHASE SHIFTS USING QUANTUM INVERSION \\[5mm]
M. SANDER, C. BECK, B.C. SCHROEDER, H.--B. PYO\\ H. BECKER,
J. BURROWS,  H.V. VON GERAMB \\
{\it Theoretische Kernphysik, Universit\"at Hamburg}
\\ {\it Luruper Chaussee 149, D--22761 Hamburg}\\[3mm]
and \\[1mm] Y. WU\\
{\it College of Engineering, Nihon University, Koriyama 963, Japan}
\\[1mm] and \\[1mm]
S. ISHIKAWA\\
{\it Physics Departement, Tohoku University, Sendai 980, Japan}
\end{center}

\begin{abstract} \small
NN and $\pi$N partial wave radial potentials have been generated using
the latest SM94--VPI(NN) and FA93--VPI($\pi$N) phase shifts. The
potentials are used to determine the deuteron properties and to compute
the $^3$H and $^3$He binding energies. The $\varepsilon_1$
mixing angles of SM94--VPI and NY93--Nijmegen differ significantly
and inversion potentials yield a P$_d$ of
                                        6.37\% and 5.78\%, respectively.
Under--binding of $^3$H and $^3$He is enhanced by SM94, which signals
more nonlocality and/or three--body potential effects than predicted
from Nijmegen phase shifts and  boson exchange models.
The local pion--nucleon $S_{31}$, $P_{31}$ and $P_{33}$ channel
potentials have been generated for guidance and
to obtain a quantitative impression in
r--space of what the $\pi$N FA93--VPI phase shifts imply.
\end{abstract}

\section*{{\normalsize\bf 1. Introduction and Motivation}}
The differential form of the Blankenbecler--Sugar equation and the
radial Schr\"odinger equation are used as the equations of motion to
determine the underlying potentials of elastic nucleon--nucleon and
pion--nucleon scattering from boundary conditions.
The most recent phase shifts SM94--VPI and FA93--VPI \cite{said}
are used as boundary conditions.
They determine the partial wave
S--matrix and Jost--functions. Our single and coupled channels
inversion algorithms use various forms of  Gelfand--Levitan and
Marchenko integral equations \cite{inversion93}.

The motivation of this analysis comes from differing phase shift
analyses results of the VPI and Nijmegen
groups irrespective of  good fits of almost the same two nucleon data
\cite{arndt94,stoks93}.
Within 0---300 MeV their phases
differ  by as much as one or two degrees.
The magnitude of the mixing angle $\varepsilon_1$ is
not consistent  within the two analyses.

We would like to point out that  the VPI analysis has important
quantitative implications for three--body  or nonlocality effects of
boson exchange potential models \cite{gibson94}. To show this we
present preliminary results of  $^3$H and $^3$He binding energy
calculations which show that the SM94 inversion potentials imply a
P$_d$ of 6.37\%
and an under--binding of three body systems by almost 1 MeV.

\section*{{\normalsize\bf 2. NN Potential Results}}
In recent years we followed all changes made to the VPI phase shifts
\cite{said} and generated local energy independent r--space
potentials for  partial waves with $L\leq 3$ and this
separately for {\em np}
and {\em pp} scattering. In the past we also used
inversion potentials whose phase shifts were taken from the Nijmegen
phase shift analysis and Bonn or Paris potential models.
Fig. 1 contains SM94--VPI inversion potentials which reproduce the
phases in the subthreshold domain 0--300 MeV well ($\leq 0.02$ degrees)
and thereafter follow  closely the data (--1.6 GeV)
with the implication of a repulsive core potential in all channels.
There is no noticeable qualitative difference existing to previous
results or NY93 inversion potentials.

\unitlength1.0cm
\begin{figure}[h]\centering
\begin{picture}(15,6)(0.0,0.0)\centering
\epsfig{figure=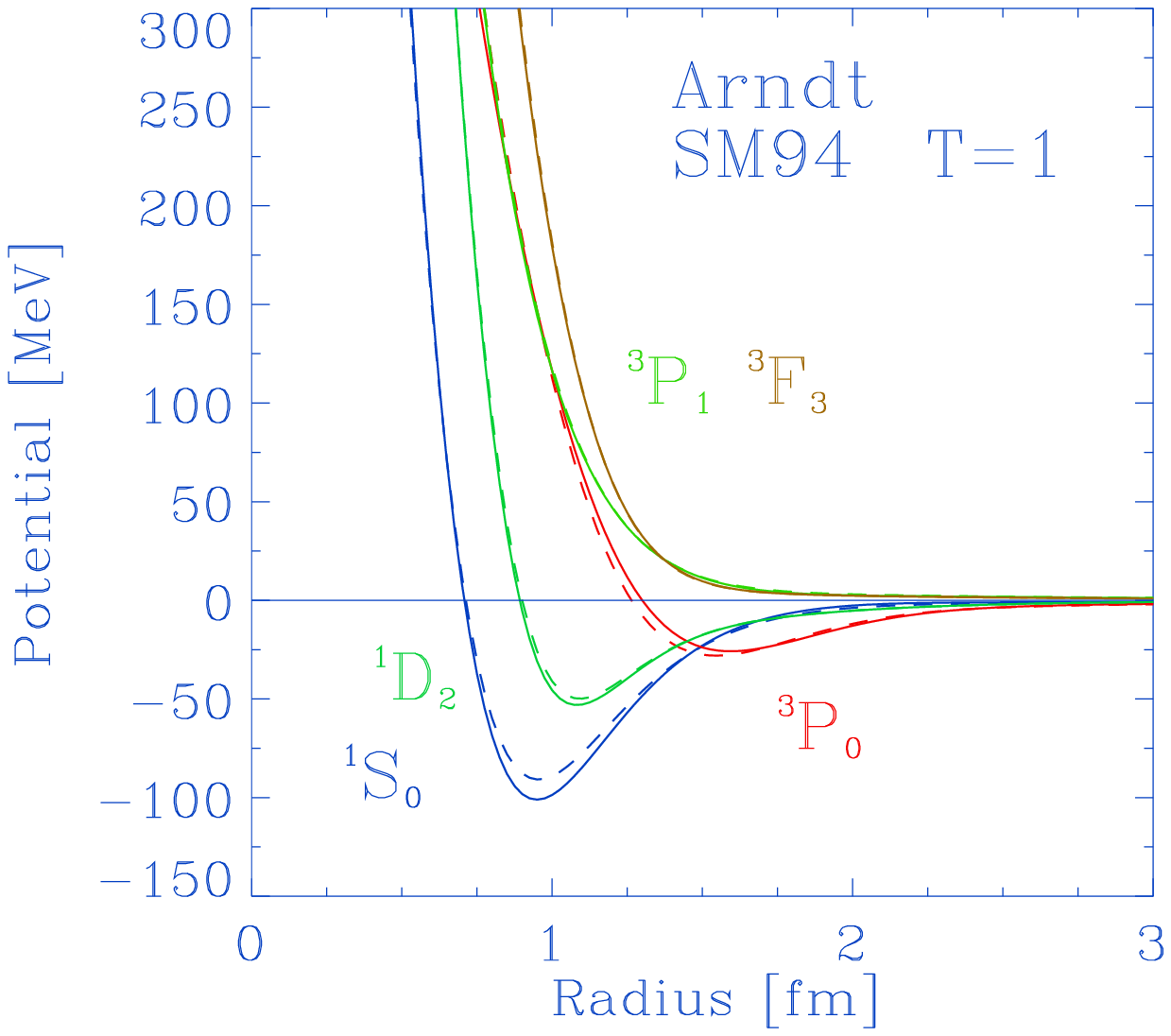,width=7.5cm}
\epsfig{figure=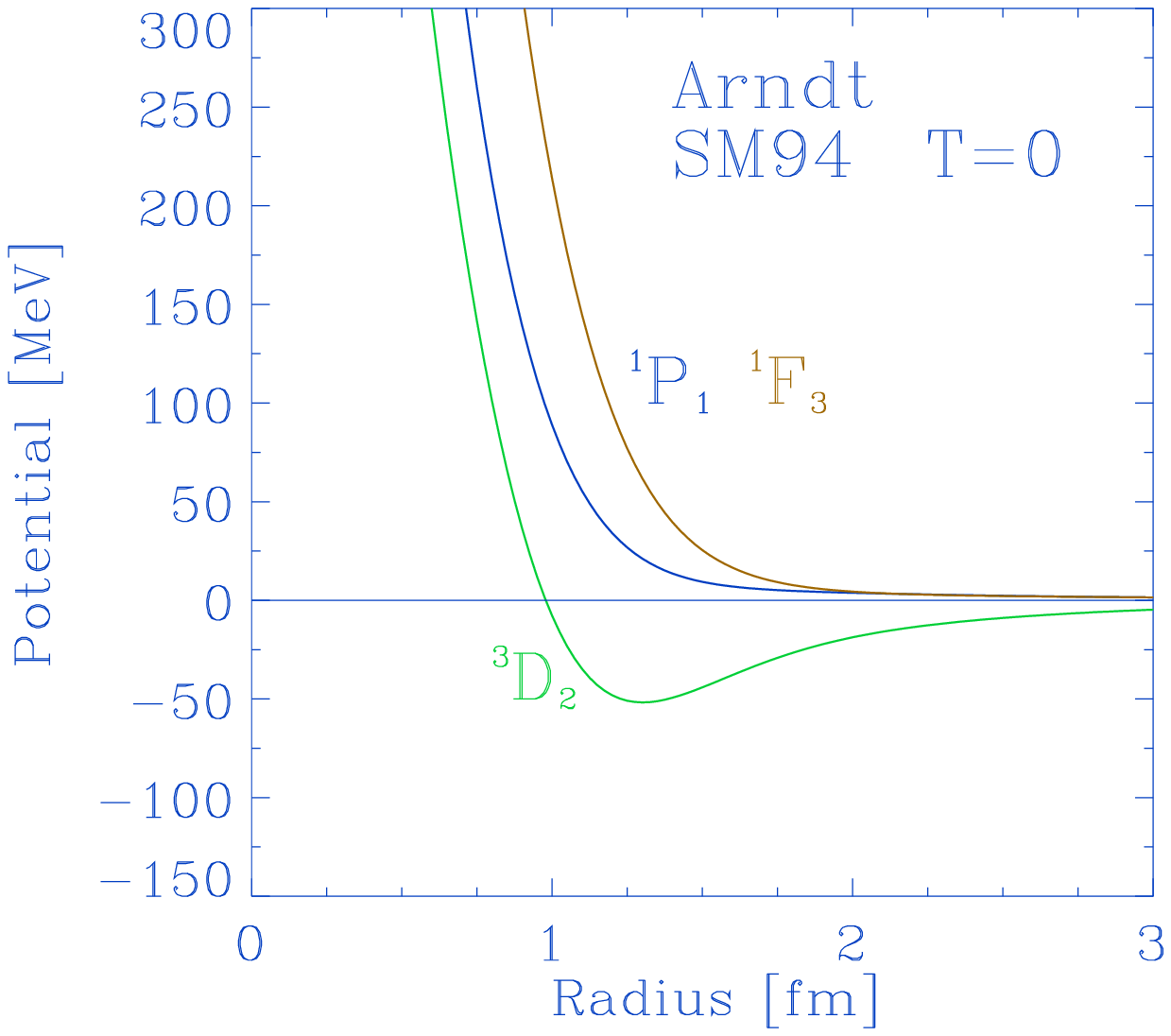,width=7.5cm}
\end{picture} \\[0.2cm]
\begin{picture}(15,5.7)(0.0,0.3)\centering
\epsfig{figure=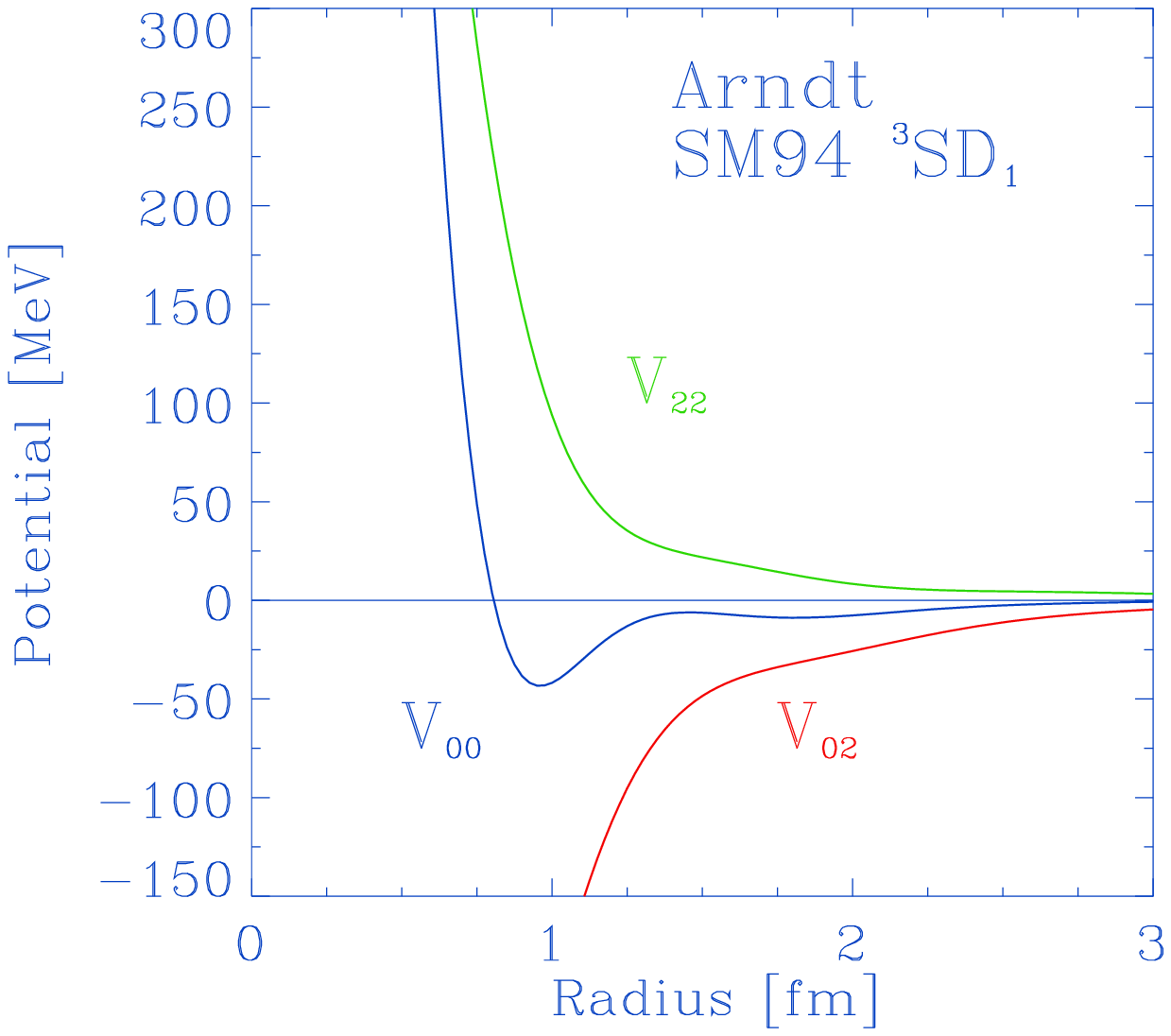,width=7.5cm}
\epsfig{figure=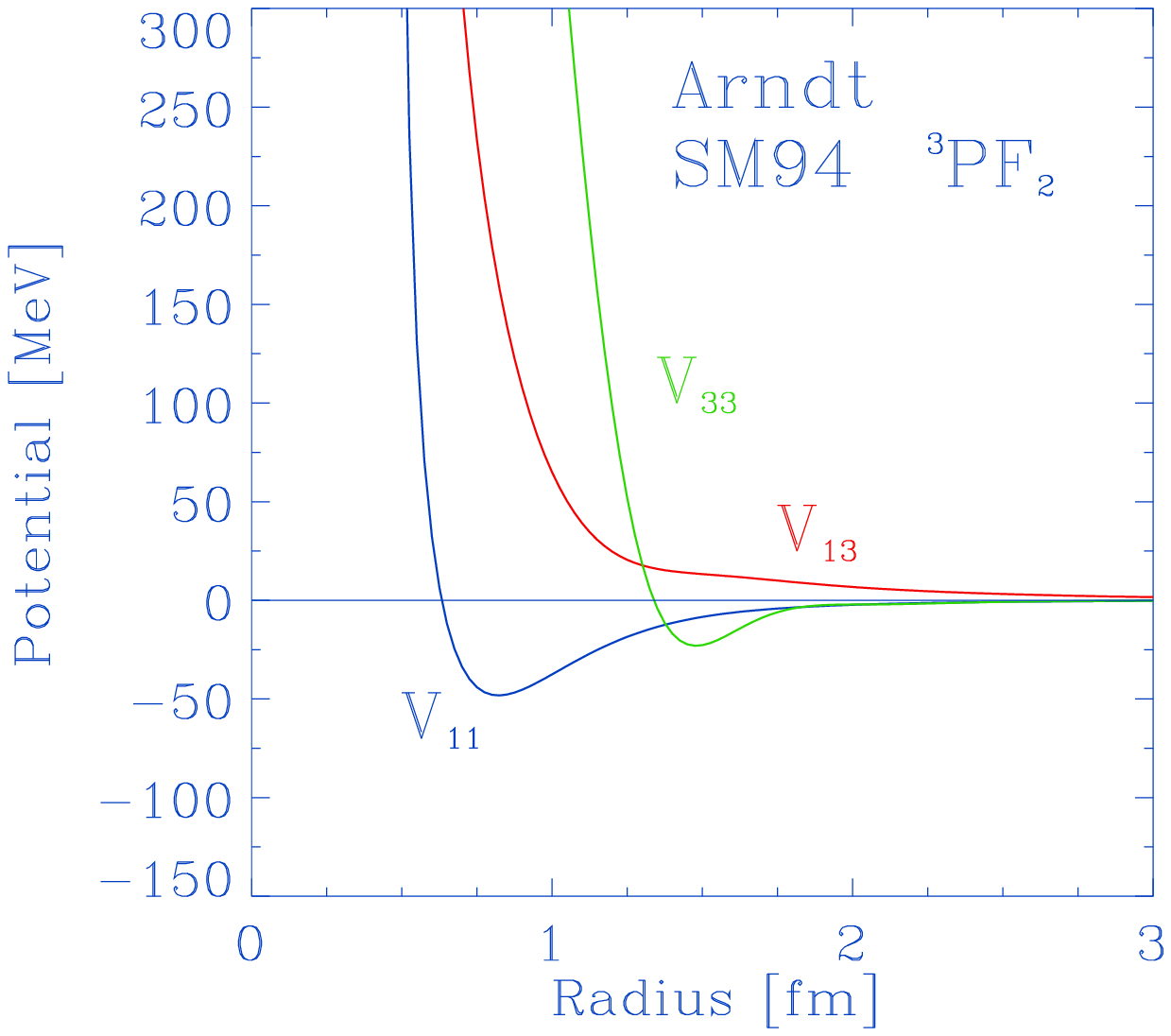,width=7.5cm}
\end{picture}
\caption{NN inversion potentials computed for SM94--VPI
phase shifts. {\em np}  (full lines)  {\em pp} (dashed lines).
All input phase shifts were retrieved from \protect\cite{said}.}
\end{figure}

In Table 1 are summarized static deuteron properties which either are
used as inversion input (E$_B$, A$_s$, $\eta$) or are derived
from inversion potentials to the quoted phase shifts. Argonne V18
results are reproduced from a recent preprint \cite{wiringa94}.

\begin{table}
\begin{center}
\caption{Deuteron  properties  from inversion
 and genuine AV18 potentials.}
\begin{tabular}[t]{ccccccc}
\hline \hline
  & SM94    &  SP94 & NY93       & AV18     & Bonn--B
           & Experiment \\
 \hline
 $E_B$ [MeV]
        & 2.22459     & 2.224638   & 2.22460    & 2.224575 & 2.224653
           & 2.22458900(22)                                \\
 $A_s$ [fm$^{1/2}$]
        & 0.8802    & 0.8860     & 0.8802     & 0.8850  & 0.8860
           & 0.8802(20)             \\
 $\eta$ & 0.0256    & 0.0263     & 0.0256     & 0.0250   & 0.0264
           & 0.0256(4)                                     \\
 $P_D$  & 6.3695   & 6.6252     & 5.7881     & 5.76     & 5.8152
           &                                               \\
 $Q$  [fm$^2$]
      & 0.276415     & 0.287457   & 0.272714   & 0.270 & 0.282740
           & 0.2859                                    \\
 $\mu$ [$\mu_0$]
       & 0.84342    & 0.84197    & 0.84673    & 0.847  & 0.84658
           & 0.857406(1)                                   \\
 $RMS$ [fm]
       & 1.962706    & 1.974389   & 1.959989   & 1.967 & 1.970987
           & 1.9650(45)                                    \\
\hline \hline
\end{tabular}
\end{center}
\end{table}

Preliminary results of three body calculations for $^3$H and $^3$He by
Y. Wu and S. Ishikawa,  using inversion potentials, are given in
Table 2. This calculation follows their recent article
\protect\cite{wu93}. Not included
in the SM94--VPI calculations is
the $^3PF_2$ coupled channel potential and thus the channels are
reduced to 20 and 24 for $J\leq 2$ and $J\leq 3$ respectively.
This  lack of potential causes the binding energies to decrease with
increasing channel numbers (a complete analysis is in progress).
Three body force effects have not been included.
As an important result of this calculation  a
large difference in  binding energy prediction
between Nijmegen (not shown Bonn-B)
and VPI appears. This is primarily
caused by the very different $\varepsilon_1$ values. To eliminate this
discrepancy we suggest a critical review of the phase shift analyses.

\begin{table}
\begin{center}
\caption{Three body  binding energies from inversion potentials.}
\begin{tabular}[b]{cccccccc}
\hline \hline
Potential     &  Channels & $^3$H  &  $^3$He &   Diff. &
$^3$H & $^3$He  &  Diff. \\
\hline
\multicolumn{5}{c}{  } & \multicolumn{3}{c}
                       {With CIB $V_{np}\neq V_{pp}$} \\
\hline
Nijm--II   &    6 & 7.834 & 7.169 &  0.665 & 7.582 & 6.931 &  0.651 \\
      &   28 & 7.852 & 7.191 &  0.661 & 7.600 & 6.952 &  0.648 \\
      &   34 & 7.849 & 7.188 &  0.661 & 7.597 & 6.949 &  0.648 \\[0.1cm]
FA91--VPI &    6 & 7.413 & 6.770 &  0.643 & 7.242 & 6.604 &  0.638 \\
      &   28 & 7.468 & 6.828 &  0.640 & 7.293 & 6.657 &  0.636 \\
      &   34 & 7.465 & 6.824 &  0.641 & 7.290 & 6.654 &  0.636 \\[0.1cm]
SM94--VPI & 6  &   7.550 & 6.894 &  0.656 & 7.225 & 6.589 &  0.636 \\
      & 20& 7.476 & 6.827 & 0.649 & 7.151 &  6.521 &   0.630 \\
      & 26& 7.471 & 6.822 & 0.649 & 7.146 &  6.616 &   0.630 \\
\hline \hline
\end{tabular}
\end{center}
\end{table}

\clearpage
\section*{{\normalsize\bf 3.  \protect$\pi$N Potential Results}}

Inversion potentials for pion--nucleon systems already have a
long standing tradition \cite{tabakin69}. Due to its importance
for medium energy physics, this topic has a very rich literature
with several excellent solutions \cite{ericson88}.

We have generated local r-space inversion potentials to a selected
sample of $\pi$N phase shifts, FA93--VPI, with the purpose
to obtain  quantitative results to the latest data in
the elastic domain. The used input
phases for $S_{31},\ P_{31}$ and $P_{33}$ are from \cite{said}.
Potential results are shown in Fig. 2. Radial dimensions,
in which  potentials significantly differ from zero, are
smaller than accustomed from NN potentials. Of particular interest
is the
$P_{33}$ resonance as shape--resonance which is localized within
the radial region 0---0.5 fm. A deep attractive well has a
radius not larger than 0.15 fm whereas the depth depends simply from
the asymptotic continuation. For future $\pi$N
work we see in inversion techniques generally
a possibility to tune  theoretical models such that they
agree perfectly with the
experiment. Also, as a didactic mean, the power of inversion procedures
cannot be denied.

\unitlength1.0cm
\begin{figure}[h]
\begin{picture}(15.2,5.9)(-0.5,0.2)\centering
\epsfig{figure=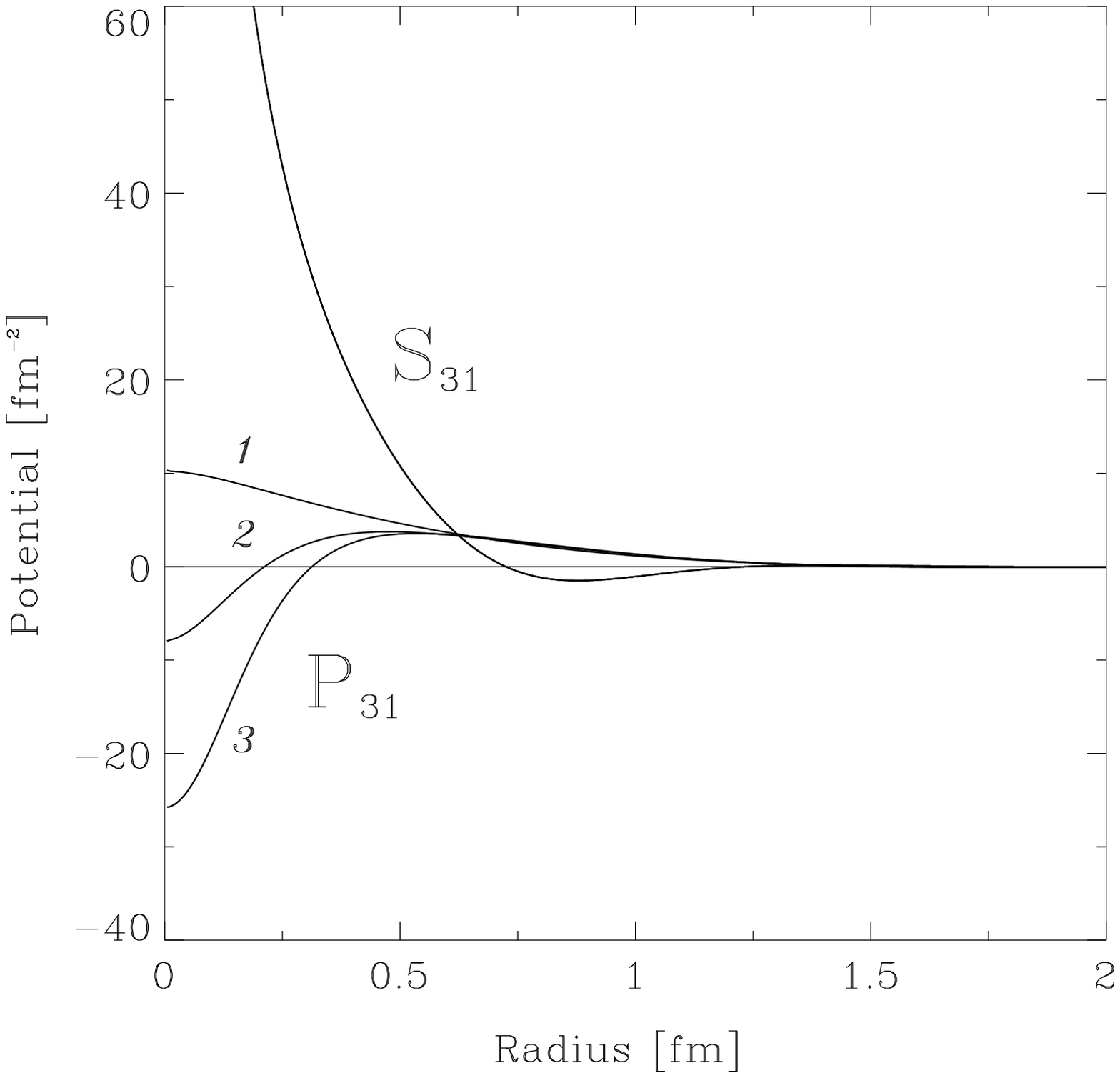,width=7.2cm}
\epsfig{figure=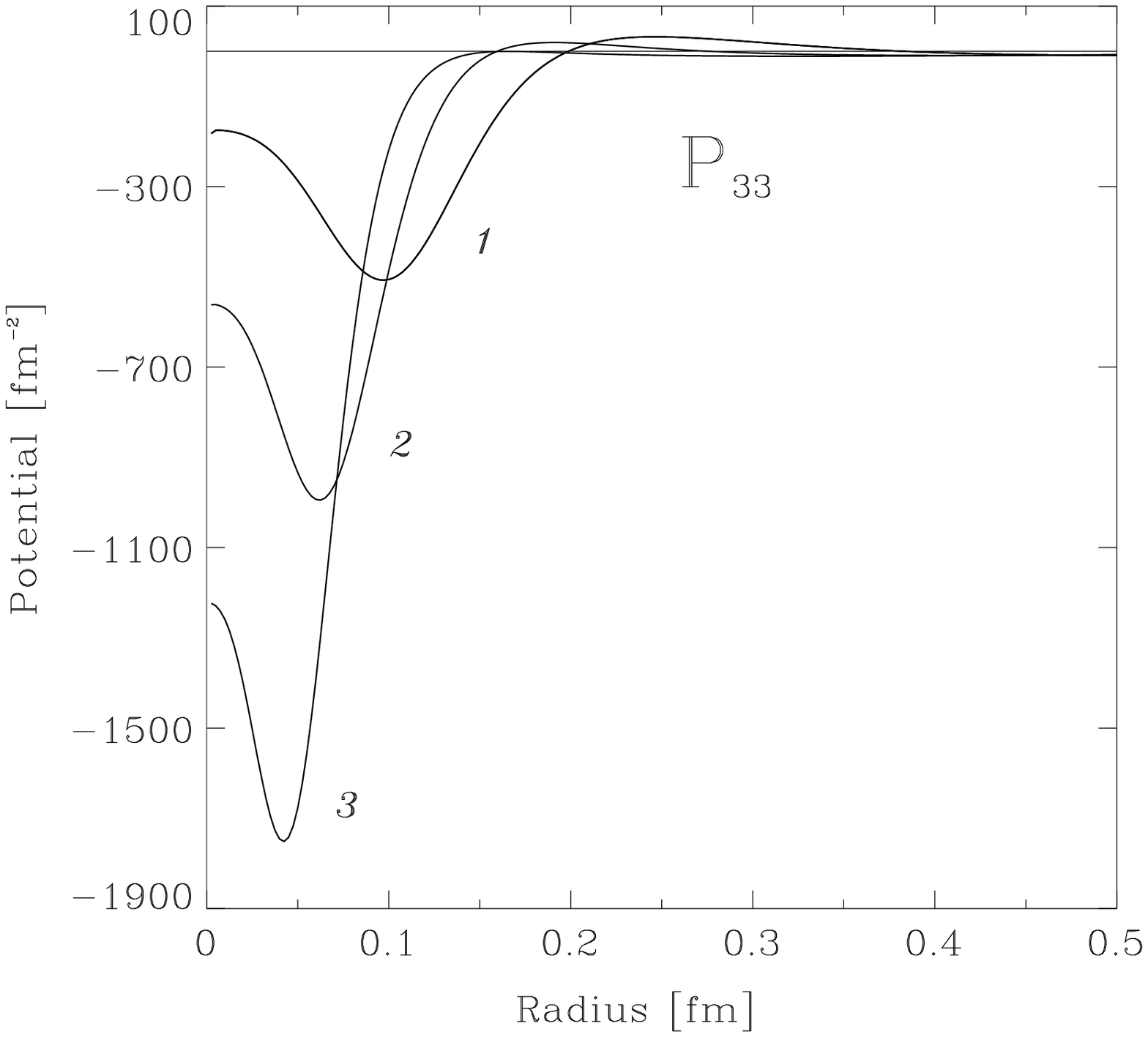,width=7.2cm}
\end{picture}
\caption{Local r--space potentials in selected channels. The
numbers 1,2 and 3 distinguish  solutions with  different  fall--offs
towards high energy, E$>$1 GeV.}
\end{figure}

\end{document}